\documentstyle[12pt,worldsci]{article}
\begin{document}
\title{
\hfill{$\mbox
{\rm JHU-TIPAC-98010}$} \\ 
\vskip 0.3in
HYBRID CHARMONIUM PRODUCTION IN NRQCD}
\author{Alexey A. Petrov\\
{\em The Johns Hopkins University, Baltimore, Maryland 21218, USA}}
\maketitle
\setlength{\baselineskip}{2.6ex}

\vspace{0.7cm}
\begin{abstract}
Using the operator product expansion and nonrelativistic QCD we study the production of charmonium hybrids in $B$ decays.  We express the decay rate in terms of a few matrix elements which eventually will be fixed by experimental measurements or calculated on the lattice.  While the Fock state expansion is problematic for hybrids, the operator product expansion still provides a model independent framework for phenomenological calculations of hybrid production and decay.  We then use a simple flux tube model to estimate the branching ratio $B\to\psi_g+X$, where $\psi_g$ is a $J^{PC}=0^{+-}$ hybrid, the large production of which could help resolve the low charm multiplicity observed in $B$ decays. 
\end{abstract}
\vspace{0.7cm}

The hybrid states remain an enigmatic feature of hadronic physics. Their existence is predicted in QCD, yet most of their properties have been studied only within models or on the lattice.  This is true even of quarkonium hybrids, because, unlike ordinary quarkonium, hybrids are characterized by excitations of the nonperturbative gluon field.  Even for large mass $m_Q$, there is no reason for a $\overline QQ$ hybrid to be a compact object, and neither the lattice nor models support such a picture. 
Contrary to the hybrids, in the limit $m_Q \to \infty$ ordinary quarkonium admits a controlled Fock state decomposition\cite{nrqcd}, in which the leading term is a $\overline QQ$ in a color singlet with fixed quantum numbers $^{2S+1}L_J$.  Higher order terms, such as color octet configurations, are suppressed by powers of $\alpha_s(m_Q)\sim v$, where $v$ is the relative velocity of the $Q$ and the $\overline Q$.  The expansion is possible because for large $m_Q$ the quarkonium is a small state, of size $(m_Qv)^{-1}$, whose interaction with the color field is governed by a multipole expansion. This velocity power counting can be made explicit by a suitable rescaling of the fields in the effective Lagrangian of NRQCD \cite{resc}. By contrast, the minimum of the hybrid meson potential is at a separation $r_0$ of order $1/\Lambda_{\rm QCD}$, independent of $m_Q$.  For large $m_Q$, fluctuations about $r_0$ are small, but the state itself is not compact. This is a generic feature of hybrid models; for example, in a constituent gluon model, the $Q$ and $\overline Q$ are in a color octet and repel each other at short distances.  At large $m_Q$, one expects a situation somewhat like a heavy rigid rotor, with nearly degenerate rotational bands of states.  This behavior has been observed in lattice studies~\cite{manke}. We can make the 
argument explicit by rescaling the coordinates in the NRQCD Lagrangian by their ``natural'' size, ${\bf x} = \lambda_x {\bf X}, ~t = \lambda_t T$, where,
contrary to the quarkonium case \cite{resc} where $\lambda_x = 1/(mv)$ we
have $\lambda_x = 1/\Lambda_{QCD}$. This rescaling explicitly takes into account the fact that in the limit $m_Q \to \infty$ hybrid quarkonium state remains finite.
A NRQCD Lagrangian can be written as
\begin{equation} \label{lagr1}
{\cal L}_{NRQCD} = i Q^\dagger ~\left[\partial^0 + i g_s
A^0 \right] Q + Q^\dagger  ~\frac{\left[{\bf \nabla} + i g_s
{\bf A} \right]^2}{2 m_Q}  Q + ~...
\end{equation}
It implies that $\partial^0$ is of the same order as $\nabla^2/2m_Q$, leading to $\lambda_t=m_Q \lambda_x^2=m_Q/\Lambda_{QCD}^2$. The fields in (\ref{lagr1}) 
should also be rescaled as $Q=\lambda_Q \Psi,~A^0=\lambda_{A^0} {\cal A}^0,~
{\bf A}=\lambda_A {\cal \bf A}$ where from the fact that 
the change of integration variables brings another factor of 
$\lambda_x^3 \lambda_t$ we obtain that $\lambda_Q^2 = \lambda_x^3$. The behavior 
of the gauge part of the Lagrangian $\frac{1}{4}G_{\mu \nu}G^{\mu \nu}$ is slightly more subtle, but it is easy to convince yourself that in the heavy quark limit
$m_Q/\Lambda_{QCD} \to \infty$ the terms $({\bf \nabla \times A})^2 \to
m_Q \lambda_x^3 \lambda_{A} ({\bf \nabla \times {\cal A}})^2$ and
$(\nabla A^0)^2 \to m_Q \lambda_x^3 \lambda_{A^0} 
(\nabla {\cal A}^0)^2$ are the leading ones with
$\lambda_{A} = \lambda_{A^0} = \sqrt{(m_Q \lambda_x^3)^{-1}} = 
\Lambda_{QCD} \sqrt{\Lambda_{QCD}/m_Q}$. This gives the rescaled NRQCD
Lagrangian for hybrid quarkonia
\begin{equation} \label{lagr2}
{\cal L}_{NRQCD}^R = i \Psi^\dagger ~\left[\partial^0 + i g_s
\sqrt{\frac{m_Q}{\Lambda_{QCD}}} {\cal A}^0 \right] \Psi + 
\frac{1}{2} \Psi^\dagger  ~\left[{\bf \nabla} + i g_s 
\sqrt{\frac{\Lambda_{QCD}}{m_Q}} {\bf \cal A} \right]^2 \Psi + ~...
\end{equation}
Clearly, all the fields and derivatives in (\ref{lagr2}) are dimensionless, so all suppression factors explicitly appear in front of the relevant operators
providing a power counting scheme of the theory. Note that Eq.(\ref{lagr2}) implies that the Coulomb (i.e. ${\cal A}^0$) gluon interactions must be resummed to all orders. Hereafter we shall use conventional fields and derivatives keeping Eq.(\ref{lagr2}) in mind to maintain correct power counting of the relevant operators.

The absence of a controlled Fock state expansion for hybrid quarkonium can be understood as follows. For the ordinary heavy quarkonium, a soft glue configuration has a small overlap with the compact two-quark
state, since the Compton wavelength of the soft gluons, of the order of $\Lambda_{QCD}^{-1}$, is much larger than the distance between the quarks. By contrast, the size of the heavy hybrid state is finite, of order $\Lambda_{QCD}^{-1}$ in the heavy quark limit, allowing for a significant interaction with the soft gluonic modes. 

Nevertheless, there is still a sort of Fock state expansion for quarkonium hybrids.  Although the gluonic degrees of freedom are intrinsically nonperturbative, we can label the various components according to the quantum numbers of the $Q\overline Q$.  For example, for the $0^{+-}$ charmonium hybrid state, we can write a decomposition of the form
\begin{equation} \label{fock}
  | \psi_g \rangle = A | c \bar c(^3S_1)_{1,8} + g_1 \rangle +
  B | c \bar c(^3P_J)_{1,8} + g_2 \rangle +
  C | c \bar c(^1S_0)_{1,8} + g_3 \rangle + \dots\,,
\end{equation}
where the subscripts $1,8$ indicate the color of the $\bar cc$ pair and the $g_i$ are various gluonic configurations.  While one is always free to do this, here there is no model-independent hierarchy among the coefficients.  This is related to the lack of a unique $\bar cc$ ``baseline'' state, and the lack of a multipole expansion to govern transitions between the different Fock components. Nonetheless, {\it within particular models\/} it is typically the case that a given hybrid is dominated by one or two $\bar cc$ configurations. Hence it is useful to organize our calculation of hybrid production in $B$ decays with a decomposition such as (\ref{fock}).
 
As an application of the formalism of nonrelativistic QCD to the physics of quarkonium hybrids, we calculate hybrid charmonium production in $B$ decays~\cite{we}. As usual, we first calculate the production rate in the framework of full QCD, separating hard and soft degrees of freedom.  Then we compute the rate in NRQCD, and use the perturbative calculation to match the coefficients of the NRQCD matrix elements.
The inclusive $B$ meson decay rate to a hybrid charmonium state $B\to\psi_g+X$ is
\begin{eqnarray} \label{gqcd}
  \Gamma^{\rm QCD} = \frac{G_F^2 \left | V_{cb} V_{cs}^* \right |^2 C_2^2}{6 m_b}
  \int \frac{d^4 P}{(2 \pi)^4}
{\rm Im} \,
  i \int d^4 y\, e^{i P y}\, \langle B | T\,\{j^{a\dagger}_\mu(y), j^b_\nu(0)
  \} |B \rangle ~
C_{ab}^{\mu \nu}, \nonumber \\
  C_{ab}^{\mu \nu} = \langle 0 |\bar c \Gamma^\mu T^a c |\psi_g + X_2
   \rangle
  \langle \psi_g + X_2 |\bar c \Gamma^\nu T^b c|0 \rangle =
  \langle 0 |\bar c \Gamma^\mu T^a c\, P_{\bar cc}\, \bar c \Gamma^\nu T^b c
  |0\rangle,
\end{eqnarray}
where $P_{\bar cc}=|\bar cc+X_3\rangle\,\langle\bar cc+X_3|$ is a projection operator onto the $\bar cc$ state in the $\psi_g$, $j_\mu^a = \bar s\Gamma_\mu T^a b$. The Eq.~(\ref{gqcd}) is then matched with the equivalent expression in NRQCD,
\begin{equation} \label{gnrqcd}
  \Gamma^{\rm NRQCD} = \frac{G_F^2 \left | V_{cb} V_{cs}^* \right |^2 C_2^2}{m_b}
  \int \frac{d^4 P}{(2 \pi)^4}
  \sum_n \frac{C_n}{m_c^{d_n-4}}\, \langle O_n^{\psi_g} \rangle\,
  2\pi\delta(P^2-m_\psi^2),
\end{equation}
where $O_n^{\psi_g}$ represents a set of NRQCD operators of increasing dimension $d_n$  
\begin{eqnarray}\label{operators}
  \langle O_8^{\psi_g}(^3S_1) \rangle &=&
  \langle 0| \chi^\dagger {\bf \sigma}^j T^a \psi P_{\psi_g}
  \psi^\dagger {\bf \sigma}^j T^a \chi |0 \rangle,~~
  \langle O_8^{\psi_g}(^1S_0) \rangle =
  \langle 0| \chi^\dagger T^a \psi P_{\psi_g}
  \psi^\dagger T^a \chi |0 \rangle,
  \nonumber \\
  \langle O_8^{\psi_g}(^3P_1) \rangle &=&
  \frac{1}{2} \langle 0| \chi^\dagger
  ( -\textstyle{i\over2}{\bf D}  {\bf \times \sigma})^j T^a \psi
  P_{\psi_g} \psi^\dagger
  ( -\textstyle{i\over2}{\bf D}  {\bf \times \sigma})^j T^a \chi 
  |0\rangle,
  \nonumber\\
  \langle P_8^{\psi_g}(^3S_1) \rangle &=&
  \frac{1}{2} \left [
  \langle 0| \chi^\dagger {\bf \sigma}^j
  ( -\textstyle{i\over2} {\bf D})^2 T^a \psi P_{\psi_g}
  \psi^\dagger {\bf \sigma}^j T^a \chi |0 \rangle
  +\langle 0|\chi^\dagger {\bf \sigma}^j T^a \psi P_{\psi_g}
  \psi^\dagger {\bf \sigma}^j
  ( -\textstyle{i\over2}{\bf D} )^2 T^a \chi|0 \rangle \right ],
  \nonumber \\
  \langle P_8^{\psi_g}(^1S_0) \rangle &=&
  \frac{1}{2} \left[
  \langle 0| \chi^\dagger
  ( -\textstyle{i\over2}{\bf D}  )^2 T^a \psi P_{\psi_g}
  \psi^\dagger T^a \chi |0 \rangle +
  \langle 0| \chi^\dagger T^a \psi P_{\psi_g}
  \psi^\dagger
  ( -\textstyle{i\over2}{\bf D}  )^2 T^a \chi |0 \rangle
  \right ],
  \nonumber \\
  \langle Q_8^{\psi_g}(^3S_1) \rangle &=&
  \frac{1}{2} \left[
  \langle 0| \chi^\dagger
  ( -\textstyle{i\over2}{\bf D}  )^j
  ( -\textstyle{i\over2}{\bf D}\cdot\sigma  )T^a \psi P_{\psi_g}
  \psi^\dagger \sigma^j T^a \chi |0 \rangle \right.\nonumber\\
  &&\qquad\qquad\left. 
  \mbox{}+\langle 0| \chi^\dagger \sigma^jT^a \psi P_{\psi_g}
  \psi^\dagger
  ( -\textstyle{i\over2}{\bf D}  )^j
  ( -\textstyle{i\over2}{\bf D}\cdot\sigma  ) T^a \chi |0 \rangle
  \right ].
\end{eqnarray}
Here $P_{\psi_g}$ is a projection onto the hybrid state $|\psi_g\rangle$ which is normalized nonrelativistically. Note that each derivative insertion brings a power of the velocity of the heavy quark, $p/m_c \sim v$:  $O_8^{\psi_g}(^3P_1)$ is suppressed compared to $O_8^{\psi_g}(^3S_1)$ by $v^2$.
We are unable to calculate matrix elements of these operators in a model independent fashion, so as usual they are left as free parameters. Since the same matrix elements would govern the total annihilation width of the hybrid or the hybrid photoproduction cross section, one should extract the leading matrix elements from a small number of experiments.

Performing the perturbative calculation of $C_{ab}^{\mu \nu}$, boosting
the relativistic four-component spinors $u_c$ to the $\bar cc$ center of mass frame and then replacing them by nonrelativistic two-component spinors $\xi$ and $\eta$
we arrive at a simple expression 
\begin{eqnarray} \label{finalqcd}
  \Gamma^{\rm QCD} &=& {{G_F^2 \left | V_{cb} V_{cs}^* \right |^2 C_2^2}\over m_b}
  \int \frac{d^4 P}{(2 \pi)^4}\frac{2 \pi m_b^2}{9\mu_\psi}
  (1- \mu_\psi)(1+2 \mu_\psi)\delta((p_b-P)^2)\times 4m_c^2\nonumber \\
  &\times& \Biggl\{
  \left (1 + \frac{{\bf q'}^2+{\bf q}^2}
  {2 m_c^2} \right )
  {\xi'}^\dagger {\bf \sigma}^j T^a \eta' 
  \eta^\dagger {\bf \sigma}^j T^a \xi+
  \frac{1}{m_c^2} {\xi'}^\dagger ({\bf q' \times \sigma})^j T^a \eta'
  \eta^\dagger ({\bf q \times \sigma})^j T^a \xi   \nonumber\\
  &&\qquad\mbox{}+
  \frac{3 \mu_\psi}{4 \mu_c (1+2 \mu_\psi)}
  \left(1 - \frac{{\bf q'}^2+{\bf q}^2}
  {2 m_c^2} \right)
  {\xi'}^\dagger T^a \eta'\eta^\dagger T^a \xi\\
  &&\mbox{}-{1\over2m_c^2}\left [
  {\xi'}^\dagger {\bf q'}^j ({\bf q' \cdot \sigma})  T^a \eta'
  \eta^\dagger \sigma^j T^a \xi +
  {\xi'}^\dagger \sigma^j T^a \eta'
  \eta^\dagger {\bf q}^j ({\bf q \cdot \sigma}) T^a \xi \right ]
  + \dots\Biggr\}\,,\nonumber
\end{eqnarray}
where $\mu_c=m_c^2/m_b^2$ and $\mu_\psi=m_\psi^2/m_b^2$.  Eq.~(\ref{finalqcd}) expresses the decay rate in terms of matrix elements of operators of increasing dimension. 

To perform the matching, we have to calculate $\Gamma (B \to \psi_g + X)$ in perturbative NRQCD using Eqs.(\ref{gnrqcd}) and (\ref{operators}) and then evaluate the NRQCD coefficients by matching full NRQCD states to perturbative ones. The partial decay rate for $B\to \psi_g+X$ is
\begin{eqnarray} \label{answer}
  \Gamma &=&\Gamma_0\,
  {4(1-\mu_\psi)^2(1 + 2 \mu_\psi )\over m_b^2 m_\psi}
   \Biggl [\langle O_8^{\psi_g}(^3S_1) \rangle
  + \frac{3 \mu_\psi}{4 \mu_c (1+2 \mu_\psi)}
  \langle O_8^{\psi_g}(^1S_0) \rangle
  +{2\over m_c^2} \langle O_8^{\psi_g}(^3P_1) \rangle \nonumber\\
  &&\quad\mbox{}
  +{1\over m_c^2} \langle P_8^{\psi_g}(^3S_1) \rangle
  -\frac{3 \mu_\psi}{4 \mu_c (1+2 \mu_\psi)m_c^2}
  \langle P_8^{\psi_g}(^1S_0) \rangle
  -{1\over m_c^2} \langle Q_8^{\psi_g}(^3S_1) \rangle+\dots
  \Biggr ].
\end{eqnarray}
where we have defined $\Gamma_0 =
C_2^2|V_{cb} V_{cs}^*|^2 G_F^2m_b^5/144\pi$. Since the Fock space decomposition of hybrid states is not controlled by a multipole expansion, the number of matrix elements important for production processes can be reduced by keeping those that are leading in powers of $1/m_c$. The formula (\ref{answer}) determines the decay rate of $B$ into a given hybrid charmonium state in terms of {\it universal\/} parameters which must be fixed by other experiments. Using the flux tube model~\cite{flux} to estimate the matrix elements (\ref{operators}) we find\cite{we}
\begin{equation}
  \Gamma(B\to\psi_g+X)\le(1\times 10^{-3})\,\Gamma(b\to c\bar cs)\,.
\end{equation}
Thus we estimate that the production of this hybrid charmonium is more than a factor of ten lower than that of typical ordinary charmonium states. It is expected that the produced hybrids, although heavy, would not primarily decay into the final states containing charmed quarks\cite{id}, thus offering a solution to the problem of low charm multiplicity in $B$ decays. However, given the tiny overall magnitude of the result, we conclude that it is quite unlikely that this mechanism could play an important role in ``hiding'' the charm produced in $b\to c\bar cs$ transitions.  It however places the studies of hybrid charmonia well within the reach of the incoming $B$ factories\cite{ph}.

\section*{Acknowledgments}
This work was supported in part by the United States National Science  Foundation under Grants No.~PHY-9404057 and PHY-9457916 and by the United States Department of Energy under Grant No.~DE-FG02-94ER40869.

\end{document}